\documentclass[reprint,aps,prl,nofootinbib]{revtex4-1}
\usepackage{graphicx}
\usepackage[utf8]{inputenc}
\usepackage{amsmath}
\usepackage{natbib}
\usepackage{bm}
\usepackage{siunitx}
\usepackage{placeins}
\usepackage[caption=false]{subfig}
\usepackage{amsmath}
\usepackage{color}
\graphicspath{{./images/}}
\usepackage{textcomp}
\usepackage{array}
\newcolumntype{P}[1]{>{\centering\arraybackslash}p{#1}}
\newcolumntype{M}[1]{>{\centering\arraybackslash}m{#1}}

\begin{document}
\title{Impact of magnetic moment and anisotropy of Co$_\textrm{1-x}$Fe$_\textrm{x}$ thin films on the magnetic proximity effect of Pt}
\author{Panagiota Bougiatioti$^1$, Orestis Manos$^1$, Olga Kuschel$^{2}$, Joachim Wollschl\"ager$^{2}$, Martin Tolkiehn$^{3}$, Sonia Francoual$^{3}$, and Timo~Kuschel$^{1}$ \email{Electronic mail: pbougiatioti@physik.uni-bielefeld.de}}
\affiliation{$^1$\mbox{Center for Spinelectronic Materials and Devices, Department of Physics,}\\
\mbox{Bielefeld University, Universit\"atsstra\ss e 25, 33615 Bielefeld, Germany}\\
$^2$\mbox{Department of Physics and Center of Physics and Chemistry of New Materials,} \\
\mbox{Osnabrück University, Barbarastra\ss e 7, 49076 Osnabrück, Germany}\\
$^3$\mbox{Deutsches Elektronen-Synchrotron DESY, Notkestraße 85, 22607 Hamburg, Germany}\\}

\date{\today}


\begin{abstract}

We present a systematic study of the magnetic proximity effect in Pt, depending on the magnetic moment and anisotropy of adjacent metallic ferromagnets. Element-selective x-ray resonant magnetic reflectivity measurements at the Pt absorption edge (11565$\,$eV) are carried out to investigate the spin polarization of Pt in Pt/Co$_\textrm{1-x}$Fe$_\textrm{x}$ bilayers. We observe the largest magnetic moment of (0.72$\,\pm\,$0.03)$\, \mu_\textrm{B}$ per spin polarized Pt atom in Pt/Co$_\textrm{33}$Fe$_\textrm{67}$, following the Slater-Pauling curve of magnetic moments in Co-Fe alloys. In general, a clear linear dependence is observed between the Pt moment and the moment of the adjacent ferromagnet. Further, we study the magnetic anisotropy of the magnetized Pt which clearly adopts the magnetic anisotropy of the ferromagnet below. This is depicted for Pt on Fe(001) and on Co$_\textrm{50}$Fe$_\textrm{50}$(001), which have a 45$^{\circ}$ relative rotation of the fourfold magnetocrystalline anisotropy.

\end{abstract}

\maketitle

\begin{figure}[!ht]
    \centering
    \includegraphics[height=6.5cm, width=\linewidth]{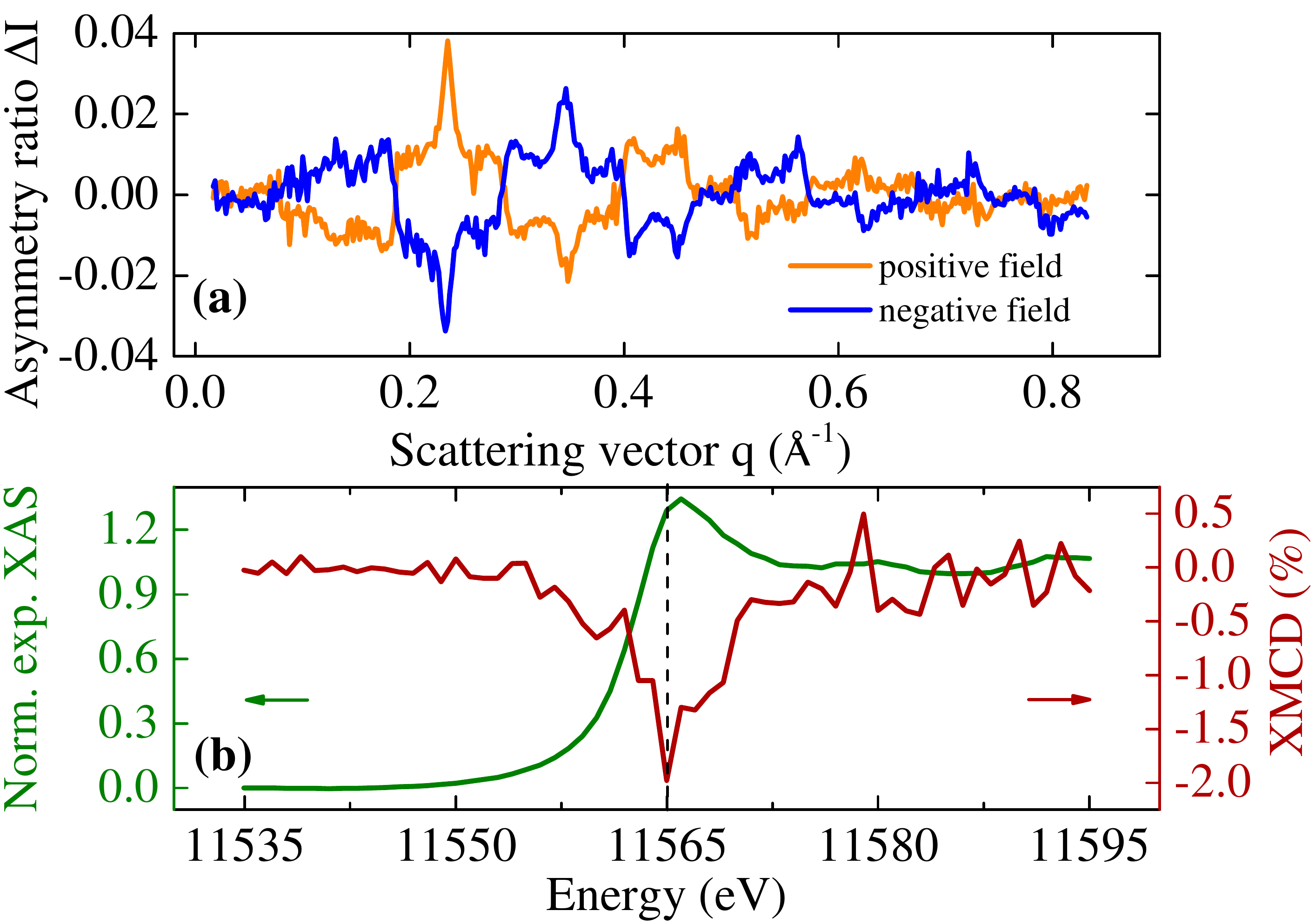}
    \caption{(a) Resonant (11565$\,$eV) asymmetry ratio $\Delta I$($q$) for both magnetic field directions. (b) Experimental energy dependant XAS spectrum (green line) and the XMCD signal (red line). All data correspond to the Pt/Co$_\textrm{50}$Fe$_\textrm{50}$ bilayer.}
    \label{fig:XRMR_XAS_XMCD}
\end{figure}

The spin polarization of a nominally paramagnetic material generated by the exchange interaction of an adjacent ferro- or ferrimagnetic material, is called magnetic proximity effect (MPE). It is most famous in Pt, which is almost ferromagnetic following the Stoner criterion description \cite{Stoner:1938}. The MPE is a key element in spintronics \cite{Hoffmann:2015} and spin caloritronics \cite{Bauer:2012}, since additional effects can contribute when Pt is employed as a spin current detector. For example, an additional proximity-induced anomalous Nernst effect can occur in spin Seebeck effect experiments \cite{Huang:2012,Kikkawa:2013PRL,BougiatiotiPRL:2017}. Furthermore, the proximity-induced anisotropic magnetoresistance can hamper spin Hall magnetoresistance studies \cite{Althammer:2013}.\\
\mbox{ } In normal metal (NM)/ferromagnet (FM) bilayer systems, the conductivity of the FM affects the MPE in the NM, as recently systematically investigated by examining the transition from FM metal (FMM) to FM insulator (FMI), using various oxygen content in Pt/NiFe$_\textrm{2}$O$_\textrm{x}$ bilayers \cite{BougiatiotiPRL:2017}. No MPE in Pt could be found except for the metallic Pt/Ni$_\textrm{33}$Fe$_\textrm{67}$ case without any oxygen. Although the MPE varies for different material combinations (e.g. FMIs vs. FMMs), general systematic studies of the MPE dependence on material parameters within one class of material are quite rare. The two most important properties of magnetic materials are the magnetic moment and the magnetic anisotropy. Therefore, in this work we systematically investigate the MPE dependence on the FM moment and FM anisotropy in NM/FMM systems.\\
\mbox{ } One way to investigate the MPE is to use x-ray magnetic circular dichroism (XMCD), allowing to extract the absolute magnetic moment per atom of each element \cite{Schutz:1989,Schutz:1990,Maruyama:1993,Antel:1999,Wilhelm:2000,Poulopoulos:2001,Wilhelm:2003,Suzuki:2005,Lu:2013,Geprags:2012,Geprags:2012Comment,Valvidares:2016,Collet:2017,Kikkawa:2017}. A much younger technique to investigate the magnetic properties of layer systems with element- and depth-sensitivity, is x-ray resonant magnetic reflectivity (XRMR) that is based on the spin-dependent interference of light reflected from the interfaces in the system \cite{Macke:2014}. This method even detects magnetic moments at buried interfaces for thicker layers \cite{Kuschel:2015}, when XMCD is not sensitive anymore. In previous studies, we investigated the spin polarization in Pt of Pt/NiFe$_\textrm{2}$O$_x$ ($4\geq x\geq0$) bilayers \cite{Kuschel:2015,Kuschel:2016,BougiatiotiPRL:2017}, in order to evaluate the proximity-induced contributions to the inverse spin Hall and anomalous Nernst voltages, while studying the transport phenomena on the samples \cite{BougiatiotiPRL:2017,BougiatiotiJAP:2017}. In addition, we examined Pt/FMM bilayers providing information about the spatial distribution of the spin polarization of Pt, across the interface to a FMM \cite{Kuschel:2015,Kuschel:2016,Klewe:2016}.\\
\mbox{ } In this work, we investigate the induced spin polarization in Pt on top of a class of material that changes the magnetic moment and anisotropy systematically with its content. This material is  Co$_\textrm{1-x}$Fe$_\textrm{x}$ with a maximum magnetic moment for Co$_\textrm{33}$Fe$_\textrm{67}$ and differently oriented magnetic anisotropy depending on the Fe content. We use XRMR to extract the Pt magnetic moments and compare the results with the magnetic moment for the Co$_\textrm{1-x}$Fe$_\textrm{x}$ layers. In addition, we detect XRMR magnetic field loops to study the magnetic anisotropy solely in the spin-polarized Pt and to compare to magnetic field loops of the Co$_\textrm{1-x}$Fe$_\textrm{x}$ layer, collected via magnetooptic Kerr effect (MOKE). For both strength of magnetic moment and magnetic anisotropy, we find a clear correlation between the spin polarized Pt and the FMM below.\\
\mbox{ } We fabricated Pt/Co$_\textrm{1-x}$Fe$_\textrm{x}$ bilayers with x$\,$=$\,0.00, 0.15, 0.30, 0.50, 0.67, 1.00$, by dc magnetron sputter deposition on top of (001)-oriented MgO substrates at room temperature (RT). The FMM layers were prepared with and without Pt in-situ deposited on top, by covering one FMM layer with a mask. The Ar pressure during the deposition for both Co$_\textrm{1-x}$Fe$_\textrm{x}$ and Pt layers was equal to $2\cdot10^{-3}$\,mbar and the base pressure of the chamber was $3\cdot10^{-9}\,$mbar. The appropriate sputter parameters were adjusted after evaluating the x-ray fluorescence spectra to achieve the desired composition.\\
\mbox{ } The XRMR and XMCD measurements were carried out at the resonant scattering and diffraction beamline P09 of the third generation synchrotron PETRA III at DESY (Hamburg, Germany) \cite{Strempfer:2013}. A fundamental theoretical background behind XRMR includes the determination of the magnetooptic parameters $\Delta\delta$ and $\Delta\beta$ which correspond to the magnetic change of the dispersion $\delta$ and absorption $\beta$ coefficients, respectively, of the investigated material exposed to x-rays of the element's absorption energy. In our case, the x-ray reflectivity (XRR) I$_{\pm}$ for left and right circularly polarized light, respectively, was detected off resonance (11465$\,$eV) and at resonance at the L$_\textrm{3}$ absorption edge of Pt (11565$\,$eV), switching fast the helicity of incident circular polarization \cite{Strempfer:2013}. Afterwards, the XRMR asymmetry ratio $\Delta I$ = $ \frac{I_+ - I_-}{I_+ + I_-}$ was calculated and the magnetic moment per Pt atom extracted using a spin depth profile model that results in a simulated asymmetry ratio fitting the experimental data. Further details of the XRMR technique, experiment, data processing, and fitting can be found in the Supplemental Material \cite{SM} (Chap. I). The XMCD spectrum was collected using an energy dispersive silicon drift detector synchronized with the piezo-actuators underneath the phase plates, allowing for the fluorescent photons for left and right circular polarized incident light to be counted separately at every point of the scan. 
In order to investigate the magnetic anisotropy of the spin polarized Pt layer, we collected XRMR magnetic field loops for different in-plane sample orientations and a fixed scattering vector $q = \frac{4\pi}{\lambda}\sin\theta$ that corresponds to a maximum asymmetry ratio $\Delta I$ ($\lambda$ is wavelength and $\theta$ is angle of incidence).\\
\mbox{ } Figure \ref{fig:XRMR_XAS_XMCD}(a) presents the XRMR asymmetry ratio $\Delta I$ for the Pt/Co$_\textrm{50}$Fe$_\textrm{50}$ bilayer plotted against the scattering vector $q$. The effect changes sign when the magnetic field direction is reversed which confirms its magnetic origin. Figure \ref{fig:XRMR_XAS_XMCD}(b) depicts the experimental energy dependant x-ray absorption spectrum (XAS, green line) at the Pt L$_\textrm{3}$ edge normalized to the edge jump, after the subtraction of a linear background. The XMCD intensity ($I_+ - I_-$)/2 is also displayed in the figure and was extracted to identify the energy with the largest dichroism. The magnetic dichroism of the spin polarized Pt has its maximum slightly below the absorption maximum (dashed line) which is in agreement with previous findings \cite{Geissler:2001,Kuschel:2015,Klewe:2016,Kuschel:2016} and, thus, the chosen energy to collect the XRMR data was at 11565$\,$eV.

\begin{figure*}[!ht]
    \centering
    \includegraphics[height=10.5cm, width=\textwidth]{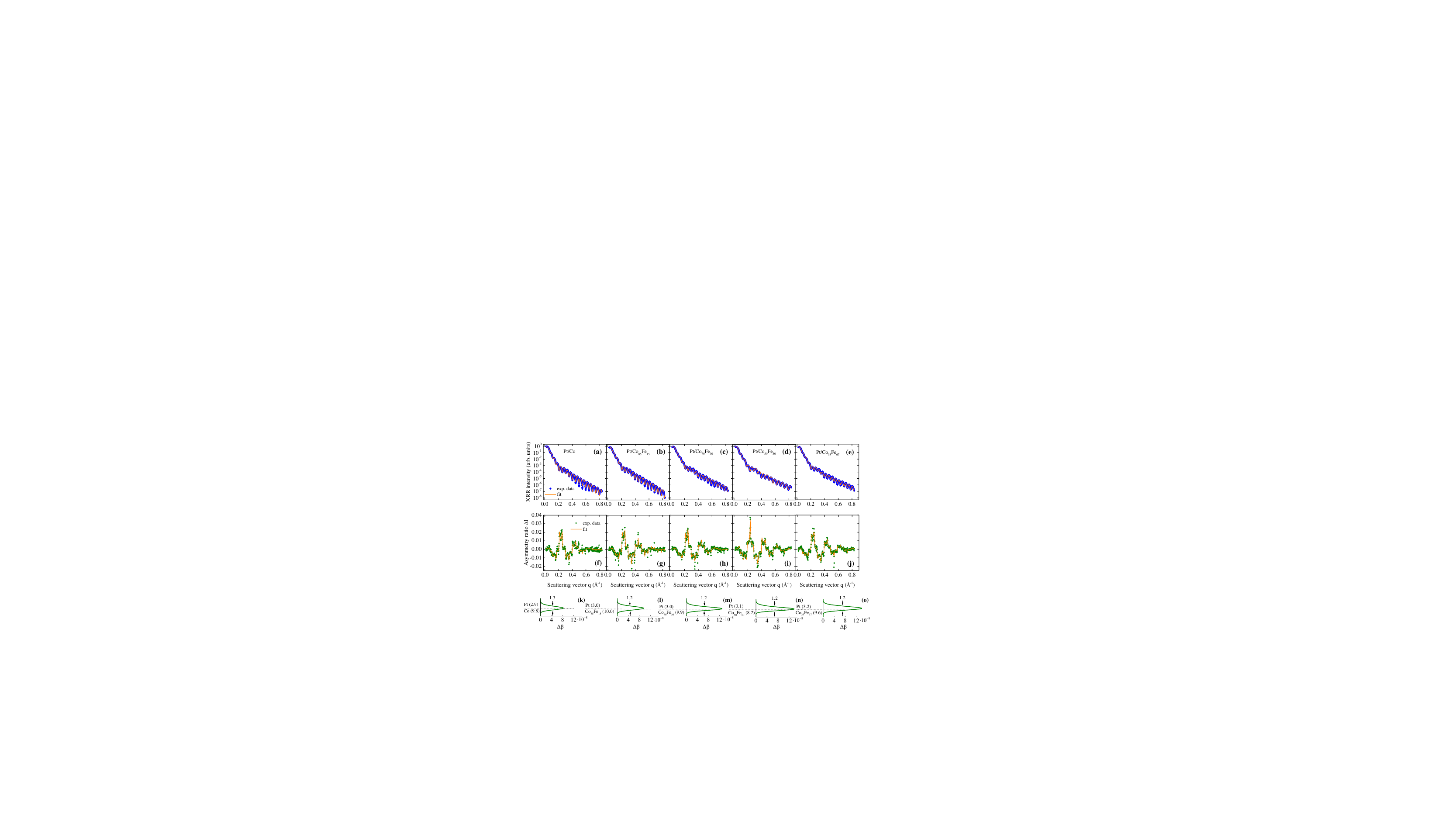}
  \caption{(a)-(e) Resonant (11565$\,$eV) XRR scans and fits with (f)-(j) the corresponding determined and fitted XRMR asymmetry ratios $\Delta I$($q$) for all Pt/Co$_\textrm{1-x}$Fe$_\textrm{x}$ bilayers. (k)-(o) Magnetooptic depth profiles which were used to fit the XRMR asymmetry ratios. The thicknesses of the layers are presented in nm. The dashed lines denote the corresponding interface position between Pt and Co$_\textrm{1-x}$Fe$_\textrm{x}$ layers. The arrows indicate the FWHM of the spin polarized layer, obtained from the structural parameters of the XRR fits.}
    \label{fig:XRMR}
\end{figure*}

Figure \ref{fig:XRMR} presents the XRR and XRMR data as well as the resulting magnetooptic depth profiles for all Pt/Co$_\textrm{1-x}$Fe$_\textrm{x}$ bilayers. Figures \ref{fig:XRMR}(a)-(e) show the averaged resonant magnetic XRR scans, collected at a photon energy of 11565$\,$eV, plotted against the scattering vector $q$ and accompanied by their fittings. Kiessig fringes appear in all scans due to the interference of the reflected light from the  Pt/Co$_\textrm{1-x}$Fe$_\textrm{x}$ and Co$_\textrm{1-x}$Fe$_\textrm{x}$/MgO interfaces. By fitting the off-resonant (11465$\,$eV) XRR curves we obtain the thickness (indicated in Figs. \ref{fig:XRMR}(k)-(o)) and roughness (typical values for the NM/FMM interfaces are between 0.20$\,$nm and 0.38$\,$nm), using literature $\beta$ and $\delta$ values for the individual layers. In a second step, we kept the structural parameters fixed for fitting the averaged resonant (11565$\,$eV) XRR curves following the description of Klewe \textit{et al}. \cite{Klewe:2016}, thus, obtaining the resonant $\beta$ and $\delta$ values. When fitting the XRMR asymmetry ratios the structural parameters from the off-resonant XRR fit and the optical values from the resonant XRR fit have been kept fixed and just the $\Delta\beta$ depth profile has been varied.\\
\mbox{ } The derived XRMR asymmetry ratios $\Delta I$($q$) are illustrated in Figs. \ref{fig:XRMR}(f)-(j), plotted together with the corresponding fittings. In all cases, pronounced oscillations are visible with an amplitude of about 2$\%$ comparable to our prior studies \cite{Kuschel:2015,Klewe:2016,Kuschel:2016,BougiatiotiPRL:2017} and additional maxima that can reach up to 4$\%$, unveiling an induced spin polarization in Pt.\\
\mbox{ } Figures \ref{fig:XRMR}(k)-(o) display the magnetooptic depth profiles of  $\Delta\beta$, which were used to fit the XRMR asymmetry ratios in Figs. \ref{fig:XRMR}(f)-(j). The magnetooptic profiles were generated by a Gaussian function at the Pt/Co$_\textrm{1-x}$Fe$_\textrm{x}$ interface, convoluted with the roughness profile of the corresponding layer \cite{Klewe:2016}. For all magnetooptic profiles, we extracted the full width at half maximum (FWHM) which represents the effective thickness of the spin polarized Pt layer at the Pt/Co$_\textrm{1-x}$Fe$_\textrm{x}$ interface. This effective spin polarized Pt thickness is between 1.2$\,$nm and 1.3$\,$nm for all samples, as indicated in Figs. \ref{fig:XRMR}(k)-(o). By comparing the experimental fit values of $\Delta\beta$ with the $\textit{ab initio}$ calculations of Ref. \cite{Kuschel:2015}, we extracted the magnetic moment per spin polarized Pt atom at the maximum of the magnetooptic profile, as summarized in Table I of the Supplemental Materials \cite{SM} (Chap. II) for all FMM compositions.

\begin{figure}[!ht]
    \centering
    \includegraphics[height=6.2cm, width=\linewidth]{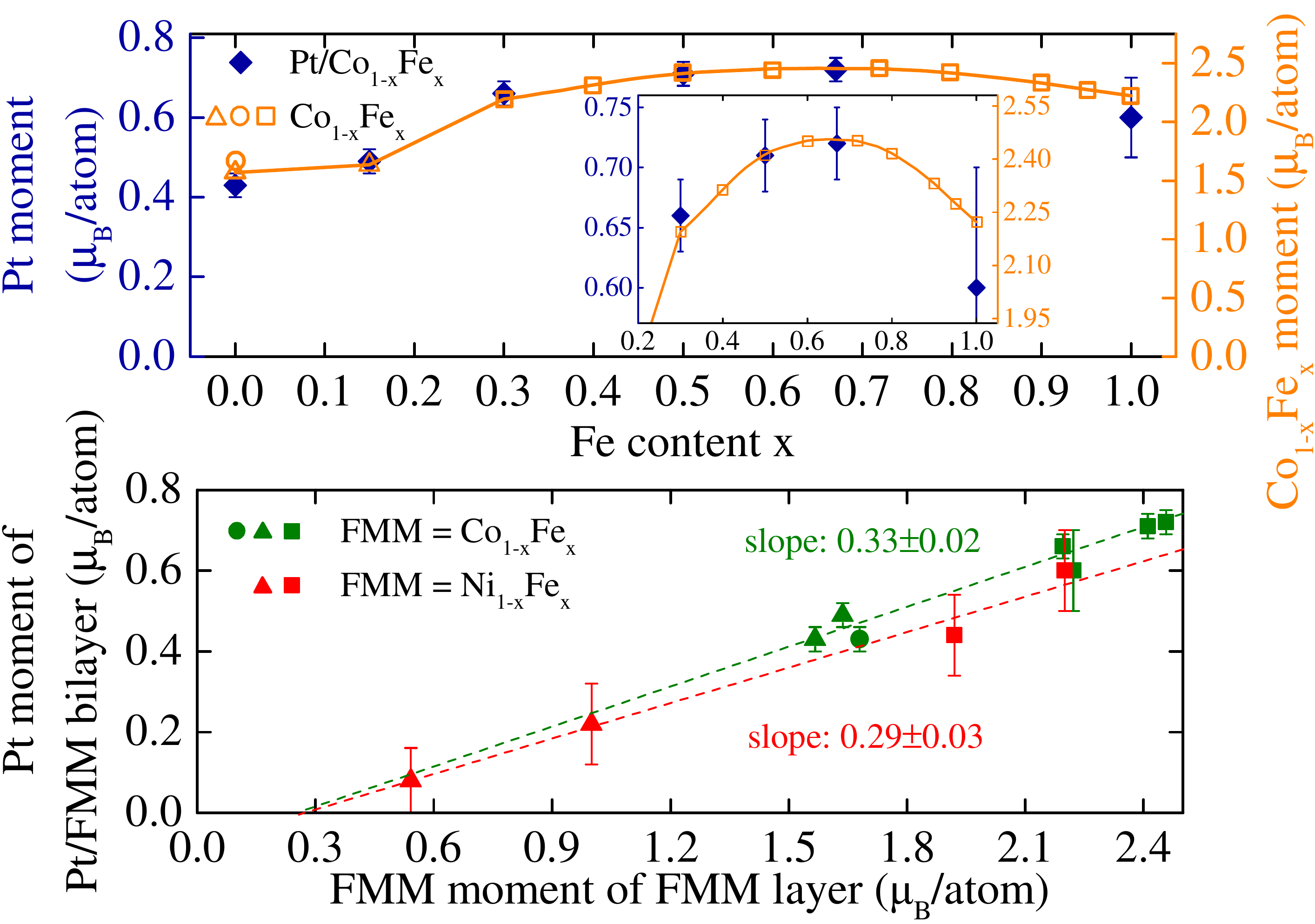}
    \caption{(a) Pt magnetic moment plotted against the Fe content x of Pt/Co$_\textrm{1-x}$Fe$_\textrm{x}$ bilayers (blue points). The orange data is taken from Ref. \cite{LandoltBornstein1986} displaying the magnetic moment per atom as derived from magnetization measurements for Co$_\textrm{1-x}$Fe$_\textrm{x}$ alloys. The inset depicts a close-up plot emphasizing the maximum experimental Pt magnetic moment in Pt/Co$_\textrm{33}$Fe$_\textrm{67}$. The data of Pt/Fe is taken from Ref. \cite{Kuschel:2015}. (b) Pt magnetic moment plotted against the FMM magnetic moment for both Pt/Co$_\textrm{1-x}$Fe$_\textrm{x}$ (green points) and Pt/Ni$_\textrm{1-x}$Fe$_\textrm{x}$ (red points, taken from Refs. \cite{Klewe:2016,Kuschel:2015}) bilayers, respectively. The green (red) dashed line is a linear fit to the data with a slope equal to $0.33\pm 0.02$ ($0.29\pm 0.03$), for the Pt/Co$_\textrm{1-x}$Fe$_\textrm{x}$ (Pt/Ni$_\textrm{1-x}$Fe$_\textrm{x}$) bilayers. In both graphs, the squares, triangles, and circles correspond to FMMs with bcc, fcc, and hcp crystal structure, respectively.}
    \label{fig:moments}
\end{figure}

Figure \ref{fig:moments}(a) presents the Pt magnetic moment for all samples (blue points), plotted against the Fe content x together with the magnetic moment values of the Co$_\textrm{1-x}$Fe$_\textrm{x}$ alloys (orange data), taken from Ref. \cite{LandoltBornstein1986}. The error bars are estimated by changing the $\beta$ values until the goodness of fit value $\chi^2$ increases up to 20$\,\%$. In our prior studies \cite{Kuschel:2015,Klewe:2016,Kuschel:2016,BougiatiotiPRL:2017} we just roughly estimated the uncertainty, therefore, the previous error values have been slightly larger. The inset depicts a close-up plot of the graph. As visible, the magnetic moments in Pt clearly exhibit a similar progress as the magnetic moments in the Co$_\textrm{1-x}$Fe$_\textrm{x}$ alloys, which follow the Slater-Pauling curve \cite{Bozorth:1951}. Both Pt and FMM moments increase with increasing x, peaking at a certain content ratio which is the Co$_\textrm{28}$Fe$_\textrm{72}$ alloy for the literature values and the Pt/Co$_\textrm{33}$Fe$_\textrm{67}$ bilayer for our experimental data. For further increase of Fe content, both Pt and FMM moments decrease. Consequently, we conclude that the strength of the magnetic coupling between the two layers depends on the magnitude of the magnetic moment in the FMM, as indicated by Klewe \textit{et al}. for Pt/Ni$_\textrm{1-x}$Fe$_\textrm{x}$ bilayers \cite{Klewe:2016} and by Poulopoulos \textit{et al}. \cite{Poulopoulos:2001} for Ni/Pt multilayers. This is valid as long as Pt is deposited on FMMs. If Pt is grown on magnetic semiconductors or insulators, the dependence of Pt moment on FM moment can be different or nonexistent due to a vanishing MPE \cite{Geprags:2012,Geprags:2012Comment,Kuschel:2015,Kuschel:2016,Valvidares:2016,Collet:2017,BougiatiotiPRL:2017}.\\
\mbox{ } Figure \ref{fig:moments}(b) exhibits the dependence of Pt magnetic moment on the FMM magnetic moment for both Pt/Co$_\textrm{1-x}$Fe$_\textrm{x}$ (green points) and Pt/Ni$_\textrm{1-x}$Fe$_\textrm{x}$ (red points, taken from Refs. \cite{Klewe:2016,Kuschel:2015}) bilayers. The dashed lines are linear fits of the data and indicate the linear dependence between the Pt and FMM magnetic moments in such bilayer systems. In addition, the slopes of both curves, as depicted in the graph, are comparable to each other considering the errors. %indicating that the induced Pt moment per FMM moment is similar for Pt on magnetic 3d metals.
The slope of the Pt moment linear dependence on the FMM moment might be interpreted as the distance to the Stoner criterion. The systematic behaviour for Pt on top of other classes of materials (such as semiconductors or slightly oxygen-reduced ferrites \cite{BougiatiotiPRL:2017}) or for other NM materials (such as Pd) on FMMs, will be part of future work.

\begin{figure}[!ht]
    \centering
    \includegraphics[height=6.5cm, width=\linewidth]{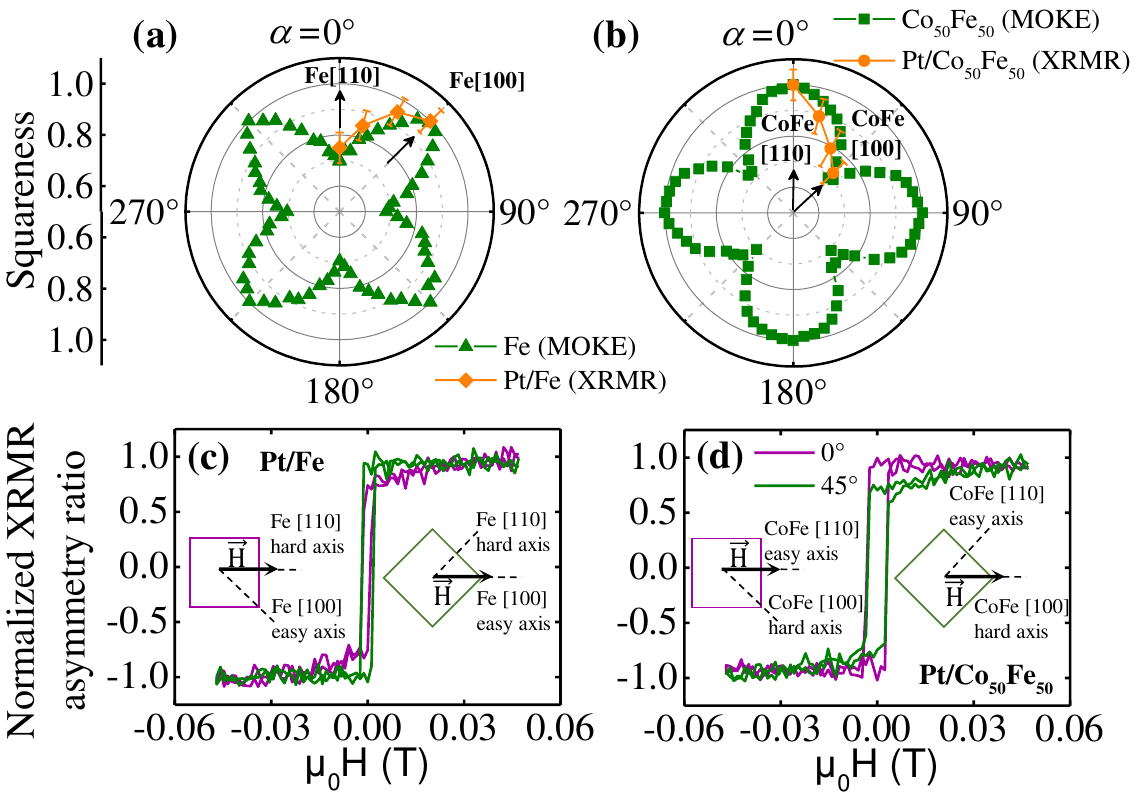}
    \caption{Squareness plotted against the value of the azimuthal angle $\alpha$ for (a) Fe (MOKE), Pt/Fe (Pt element-selective XRMR) and (b) Co$_\textrm{50}$Fe$_\textrm{50}$ (MOKE), Pt/Co$_\textrm{50}$Fe$_\textrm{50}$ (Pt element-selective XRMR) samples, collected at RT. The angle $\alpha$ corresponds to the angle between the direction of the applied magnetic field and the [110] direction of the corresponding alloy, as sketched in the inset of Figs. \ref{fig:Anisotropy}(c) and (d). The purple (green) square corresponds to the orientation of the MgO substrate with respect to the direction of the applied magnetic field leading to $\alpha=0^\circ$ ($\alpha=45^\circ$). Detected XRMR asymmetry ratio field loops with $\alpha=0^\circ,45^\circ$ for (c) Pt/Fe and (d) Pt/Co$_\textrm{50}$Fe$_\textrm{50}$ bilayers, collected in resonance (11565$\,$eV) and with a $q$ value of 0.24$\,$\r{A}$^{-1}$.}
    \label{fig:Anisotropy}
\end{figure}

As a next step, we investigated the magnetic anisotropy of Fe and Co$_\textrm{50}$Fe$_\textrm{50}$ samples by performing MOKE rotational measurements with different in-plane crystal orientation directions ($0^\circ \geq \alpha \geq 360^\circ$ in steps of $5^\circ$), in the presence of an in-plane magnetic field. The azimuthal angle $\alpha$ corresponds to the angle between the direction of the applied magnetic field and the [110] direction of the corresponding alloy, as sketched in the inset of Figs. \ref{fig:Anisotropy}(c) and (d). In order to examine the magnetic anisotropy of the spin polarized Pt layer, we collected XRMR field loop measurements, for different sample orientations ($0^\circ \geq \alpha \geq 45^\circ$ in steps of $15^\circ$).\\
\mbox{ } Figures \ref{fig:Anisotropy}(a) and (b) present the squareness which is the ratio between the magnetic remanence and the saturation magnetization, extracted from the MOKE loops (not shown) for the Fe and Co$_\textrm{50}$Fe$_\textrm{50}$ samples and from XRMR field loops for the Pt/Fe and Pt/Co$_\textrm{50}$Fe$_\textrm{50}$ samples. For the Fe film (cf. Fig. \ref{fig:Anisotropy}(a)), the MOKE measurements reveal magnetic easy axes along the Fe$<$100$>$ directions, which correspond to the MgO$<$110$>$ directions ($\alpha=45^\circ, 135^\circ, 225^\circ, 315^\circ$) with high remanence values and magnetic hard axes along the the Fe$<$110$>$ directions corresponding to the MgO$<$100$>$ directions ($\alpha=0^\circ, 90^\circ, 180^\circ, 270^\circ$) with low remanence values \cite{Costa:2000,Kuschel:2011}. On the other hand, for the Co$_\textrm{50}$Fe$_\textrm{50}$ sample (cf. Fig. \ref{fig:Anisotropy}(b)) the four-fold magnetic anisotropy is 45$^\circ$ rotated compared to Fe. Therefore, the magnetic easy axes for the Co$_\textrm{50}$Fe$_\textrm{50}$ sample are aligned along the CoFe$<$110$>$ directions which are the MgO$<$100$>$ directions with high remanence values, whereas the magnetic hard axes are aligned along the CoFe$<$100$>$ directions corresponding to the MgO$<$110$>$ directions with low remanence values \cite{Shikada:2009,Kuschel:2012}.\\
\mbox{ } In Figs. \ref{fig:Anisotropy}(c) and (d) two normalized XRMR field loops for Pt/Fe and Pt/Co$_\textrm{50}$Fe$_\textrm{50}$ samples are illustrated for $\alpha=0^\circ$ and $45^\circ$, respectively. For the Pt/Fe bilayer (Fig. \ref{fig:Anisotropy}(c)), the spin polarized Pt layer showed the magnetic easy axis for $\alpha=45^\circ$ (green curve) and the magnetic hard axis for $\alpha=0^\circ$ (purple curve), which is consistent with the results extracted from the previously analyzed MOKE measurements of the pure Fe layer. In addition, for the Pt/Co$_\textrm{50}$Fe$_\textrm{50}$ sample (Fig. \ref{fig:Anisotropy}(d)) the magnetic easy axis appeared for $\alpha=0^\circ$ (purple curve) and the magnetic hard axis was found at $\alpha=45^\circ$ (green curve), which also coincides with the MOKE results of the pure Co$_\textrm{50}$Fe$_\textrm{50}$ film. Moreover, the extracted squareness values from the XRMR field loops for the Pt/Fe and Pt/Co$_\textrm{50}$Fe$_\textrm{50}$ samples are included in Figs. \ref{fig:Anisotropy}(a) and (b), respectively. The obtained results clearly show that the magnetic anisotropy of the spin polarized Pt layer on top of the FMM (where FMM is Fe or Co$_\textrm{50}$Fe$_\textrm{50}$) adopts the magnetic anisotropy of the FMM.\\
\mbox{ } In conclusion, we investigated the dependence of the MPE of Pt on the magnetic moment and anisotropy of different FMMs below, performing XRMR and MOKE measurements. We fabricated Pt/Co$_\textrm{1-x}$Fe$_\textrm{x}$ bilayers by dc magnetron sputter deposition on top of (001)-oriented MgO substrates. The XRMR asymmetry ratios were quantitatively analyzed and the largest spin polarization was found in Pt/Co$_\textrm{33}$Fe$_\textrm{67}$ with a magnetic moment equal to $m_\textrm{Pt}=(0.72\,\pm\,0.03)\, \mu_\textrm{B}$ per spin polarized Pt atom. We also found that the magnetic moment in Pt follows the Slater-Pauling curve of magnetic moments in Co$_\textrm{1-x}$Fe$_\textrm{x}$ alloys. In addition, the Pt moment is increasing linearly with the FMM moment and the slope of this linear dependence is quite similar when we compare the Pt/Ni$_\textrm{1-x}$Fe$_\textrm{x}$ and Pt/Co$_\textrm{1-x}$Fe$_\textrm{x}$ bilayers. Furthermore, we examined the correlation between the magnetic anisotropy in the FMM layers (Fe and Co$_\textrm{50}$Fe$_\textrm{50}$ extracted via MOKE measurements) and the magnetic anisotropy of the spin polarized Pt layer (extracted via magnetic reflectivity measurements at the Pt L$_3$ absorption edge) of the Pt/Fe and Pt/Co$_\textrm{50}$Fe$_\textrm{50}$ bilayers. The magnetic anisotropy of the spin polarized Pt layer reflects the magnetic anisotropy of the FMM in the corresponding alloy. Thus, we can conclude, that the Pt nicely follows the FMM magnetization strength and orientation. For Pt on FMIs this is still an open question beside the general question of MPE in Pt/FMIs and will be explored in future research. Moreover, additional research should be conducted in such bilayer systems to create a full FM moment and conductivity mapping of the MPE in various NMs, to further elucidate the details of the MPE mechanism.\\
\mbox{ } The authors gratefully acknowledge financial support by the Deutsche Forschungsgemeinschaft (DFG) within the Priority Program Spin Caloric Transport (SPP 1538, KU 3271/1–1) and the Deutsche Elektronen Synchrotron (DESY). Further they are grateful for the opportunity to work at beamline P09, PETRA III at the Deutsche Elektronen Synchrotron, a member of the Helmholtz Association (HGF) and for technical support by David Reuther and J\"org Strempfer. They also thank G\"unter Reiss from Bielefeld University, Germany, for making available the laboratory equipment needed for sample preparation and characterization.

\bibliographystyle{apsrev4-1}
\bibliography{main}

\begin{thebibliography}{35}%
\makeatletter
\providecommand \@ifxundefined [1]{%
 \@ifx{#1\undefined}
}%
\providecommand \@ifnum [1]{%
 \ifnum #1\expandafter \@firstoftwo
 \else \expandafter \@secondoftwo
 \fi
}%
\providecommand \@ifx [1]{%
 \ifx #1\expandafter \@firstoftwo
 \else \expandafter \@secondoftwo
 \fi
}%
\providecommand \natexlab [1]{#1}%
\providecommand \enquote  [1]{``#1''}%
\providecommand \bibnamefont  [1]{#1}%
\providecommand \bibfnamefont [1]{#1}%
\providecommand \citenamefont [1]{#1}%
\providecommand \href@noop [0]{\@secondoftwo}%
\providecommand \href [0]{\begingroup \@sanitize@url \@href}%
\providecommand \@href[1]{\@@startlink{#1}\@@href}%
\providecommand \@@href[1]{\endgroup#1\@@endlink}%
\providecommand \@sanitize@url [0]{\catcode `\\12\catcode `\$12\catcode
  `\&12\catcode `\#12\catcode `\^12\catcode `\_12\catcode `\%12\relax}%
\providecommand \@@startlink[1]{}%
\providecommand \@@endlink[0]{}%
\providecommand \url  [0]{\begingroup\@sanitize@url \@url }%
\providecommand \@url [1]{\endgroup\@href {#1}{\urlprefix }}%
\providecommand \urlprefix  [0]{URL }%
\providecommand \Eprint [0]{\href }%
\providecommand \doibase [0]{http://dx.doi.org/}%
\providecommand \selectlanguage [0]{\@gobble}%
\providecommand \bibinfo  [0]{\@secondoftwo}%
\providecommand \bibfield  [0]{\@secondoftwo}%
\providecommand \translation [1]{[#1]}%
\providecommand \BibitemOpen [0]{}%
\providecommand \bibitemStop [0]{}%
\providecommand \bibitemNoStop [0]{.\EOS\space}%
\providecommand \EOS [0]{\spacefactor3000\relax}%
\providecommand \BibitemShut  [1]{\csname bibitem#1\endcsname}%
\let\auto@bib@innerbib\@empty
%</preamble>
\bibitem [{\citenamefont {Stoner}(1938)}]{Stoner:1938}%
  \BibitemOpen
  \bibfield  {author} {\bibinfo {author} {\bibfnamefont {E.~C.}\ \bibnamefont
  {Stoner}},\ }\href {\doibase 10.1098/rspa.1938.0066} {\bibfield  {journal}
  {\bibinfo  {journal} {Proc. R. Soc.}\ }\textbf {\bibinfo {volume} {165}},\
  \bibinfo {pages} {372} (\bibinfo {year} {1938})}\BibitemShut {NoStop}%
\bibitem [{\citenamefont {Hoffmann}\ and\ \citenamefont
  {Bader}(2015)}]{Hoffmann:2015}%
  \BibitemOpen
  \bibfield  {author} {\bibinfo {author} {\bibfnamefont {A.}~\bibnamefont
  {Hoffmann}}\ and\ \bibinfo {author} {\bibfnamefont {S.~D.}\ \bibnamefont
  {Bader}},\ }\href {\doibase 10.1103/PhysRevApplied.4.047001} {\bibfield
  {journal} {\bibinfo  {journal} {Phys. Rev. Applied}\ }\textbf {\bibinfo
  {volume} {4}},\ \bibinfo {pages} {047001} (\bibinfo {year}
  {2015})}\BibitemShut {NoStop}%
\bibitem [{\citenamefont {Bauer}\ \emph {et~al.}(2012)\citenamefont {Bauer},
  \citenamefont {Saitoh},\ and\ \citenamefont {van Wees}}]{Bauer:2012}%
  \BibitemOpen
  \bibfield  {author} {\bibinfo {author} {\bibfnamefont {G.~E.~W.}\
  \bibnamefont {Bauer}}, \bibinfo {author} {\bibfnamefont {E.}~\bibnamefont
  {Saitoh}}, \ and\ \bibinfo {author} {\bibfnamefont {B.~J.}\ \bibnamefont {van
  Wees}},\ }\href {\doibase 10.1038/nmat3301} {\bibfield  {journal} {\bibinfo
  {journal} {Nat. Mater.}\ }\textbf {\bibinfo {volume} {11}},\ \bibinfo {pages}
  {391} (\bibinfo {year} {2012})}\BibitemShut {NoStop}%
\bibitem [{\citenamefont {Huang}\ \emph {et~al.}(2012)\citenamefont {Huang},
  \citenamefont {Fan}, \citenamefont {Qu}, \citenamefont {Chen}, \citenamefont
  {Wang}, \citenamefont {Wu}, \citenamefont {Chen}, \citenamefont {Xiao},\ and\
  \citenamefont {Chien}}]{Huang:2012}%
  \BibitemOpen
  \bibfield  {author} {\bibinfo {author} {\bibfnamefont {S.~Y.}\ \bibnamefont
  {Huang}}, \bibinfo {author} {\bibfnamefont {X.}~\bibnamefont {Fan}}, \bibinfo
  {author} {\bibfnamefont {D.}~\bibnamefont {Qu}}, \bibinfo {author}
  {\bibfnamefont {Y.~P.}\ \bibnamefont {Chen}}, \bibinfo {author}
  {\bibfnamefont {W.~G.}\ \bibnamefont {Wang}}, \bibinfo {author}
  {\bibfnamefont {J.}~\bibnamefont {Wu}}, \bibinfo {author} {\bibfnamefont
  {T.~Y.}\ \bibnamefont {Chen}}, \bibinfo {author} {\bibfnamefont {J.~Q.}\
  \bibnamefont {Xiao}}, \ and\ \bibinfo {author} {\bibfnamefont {C.~L.}\
  \bibnamefont {Chien}},\ }\href {\doibase 10.1103/PhysRevLett.109.107204}
  {\bibfield  {journal} {\bibinfo  {journal} {Phys. Rev. Lett.}\ }\textbf
  {\bibinfo {volume} {109}},\ \bibinfo {pages} {107204} (\bibinfo {year}
  {2012})}\BibitemShut {NoStop}%
\bibitem [{\citenamefont {Kikkawa}\ \emph {et~al.}(2013)\citenamefont
  {Kikkawa}, \citenamefont {Uchida}, \citenamefont {Shiomi}, \citenamefont
  {Qiu}, \citenamefont {Hou}, \citenamefont {Tian}, \citenamefont {Nakayama},
  \citenamefont {Jin},\ and\ \citenamefont {Saitoh}}]{Kikkawa:2013PRL}%
  \BibitemOpen
  \bibfield  {author} {\bibinfo {author} {\bibfnamefont {T.}~\bibnamefont
  {Kikkawa}}, \bibinfo {author} {\bibfnamefont {K.}~\bibnamefont {Uchida}},
  \bibinfo {author} {\bibfnamefont {Y.}~\bibnamefont {Shiomi}}, \bibinfo
  {author} {\bibfnamefont {Z.}~\bibnamefont {Qiu}}, \bibinfo {author}
  {\bibfnamefont {D.}~\bibnamefont {Hou}}, \bibinfo {author} {\bibfnamefont
  {D.}~\bibnamefont {Tian}}, \bibinfo {author} {\bibfnamefont {H.}~\bibnamefont
  {Nakayama}}, \bibinfo {author} {\bibfnamefont {X.-F.}\ \bibnamefont {Jin}}, \
  and\ \bibinfo {author} {\bibfnamefont {E.}~\bibnamefont {Saitoh}},\ }\href
  {\doibase 10.1103/PhysRevLett.110.067207} {\bibfield  {journal} {\bibinfo
  {journal} {Phys. Rev. Lett.}\ }\textbf {\bibinfo {volume} {110}},\ \bibinfo
  {pages} {067207} (\bibinfo {year} {2013})}\BibitemShut {NoStop}%
\bibitem [{\citenamefont {Bougiatioti}\ \emph
  {et~al.}(2017{\natexlab{a}})\citenamefont {Bougiatioti}, \citenamefont
  {Klewe}, \citenamefont {Meier}, \citenamefont {Manos}, \citenamefont
  {Kuschel}, \citenamefont {Wollschl\"ager}, \citenamefont {Bouchenoire},
  \citenamefont {Brown}, \citenamefont {Schmalhorst}, \citenamefont {Reiss},\
  and\ \citenamefont {Kuschel}}]{BougiatiotiPRL:2017}%
  \BibitemOpen
  \bibfield  {author} {\bibinfo {author} {\bibfnamefont {P.}~\bibnamefont
  {Bougiatioti}}, \bibinfo {author} {\bibfnamefont {C.}~\bibnamefont {Klewe}},
  \bibinfo {author} {\bibfnamefont {D.}~\bibnamefont {Meier}}, \bibinfo
  {author} {\bibfnamefont {O.}~\bibnamefont {Manos}}, \bibinfo {author}
  {\bibfnamefont {O.}~\bibnamefont {Kuschel}}, \bibinfo {author} {\bibfnamefont
  {J.}~\bibnamefont {Wollschl\"ager}}, \bibinfo {author} {\bibfnamefont
  {L.}~\bibnamefont {Bouchenoire}}, \bibinfo {author} {\bibfnamefont {S.~D.}\
  \bibnamefont {Brown}}, \bibinfo {author} {\bibfnamefont {J.-M.}\ \bibnamefont
  {Schmalhorst}}, \bibinfo {author} {\bibfnamefont {G.}~\bibnamefont {Reiss}},
  \ and\ \bibinfo {author} {\bibfnamefont {T.}~\bibnamefont {Kuschel}},\ }\href
  {\doibase 10.1103/PhysRevLett.119.227205} {\bibfield  {journal} {\bibinfo
  {journal} {Phys. Rev. Lett.}\ }\textbf {\bibinfo {volume} {119}},\ \bibinfo
  {pages} {227205} (\bibinfo {year} {2017}{\natexlab{a}})}\BibitemShut
  {NoStop}%
\bibitem [{\citenamefont {Althammer}\ \emph {et~al.}(2013)\citenamefont
  {Althammer}, \citenamefont {Meyer}, \citenamefont {Nakayama}, \citenamefont
  {Schreier}, \citenamefont {Altmannshofer}, \citenamefont {Weiler},
  \citenamefont {Huebl}, \citenamefont {Gepr\"ags}, \citenamefont {Opel},
  \citenamefont {Gross}, \citenamefont {Meier}, \citenamefont {Klewe},
  \citenamefont {Kuschel}, \citenamefont {Schmalhorst}, \citenamefont {Reiss},
  \citenamefont {Shen}, \citenamefont {Gupta}, \citenamefont {Chen},
  \citenamefont {Bauer}, \citenamefont {Saitoh},\ and\ \citenamefont
  {Goennenwein}}]{Althammer:2013}%
  \BibitemOpen
  \bibfield  {author} {\bibinfo {author} {\bibfnamefont {M.}~\bibnamefont
  {Althammer}}, \bibinfo {author} {\bibfnamefont {S.}~\bibnamefont {Meyer}},
  \bibinfo {author} {\bibfnamefont {H.}~\bibnamefont {Nakayama}}, \bibinfo
  {author} {\bibfnamefont {M.}~\bibnamefont {Schreier}}, \bibinfo {author}
  {\bibfnamefont {S.}~\bibnamefont {Altmannshofer}}, \bibinfo {author}
  {\bibfnamefont {M.}~\bibnamefont {Weiler}}, \bibinfo {author} {\bibfnamefont
  {H.}~\bibnamefont {Huebl}}, \bibinfo {author} {\bibfnamefont
  {S.}~\bibnamefont {Gepr\"ags}}, \bibinfo {author} {\bibfnamefont
  {M.}~\bibnamefont {Opel}}, \bibinfo {author} {\bibfnamefont {R.}~\bibnamefont
  {Gross}}, \bibinfo {author} {\bibfnamefont {D.}~\bibnamefont {Meier}},
  \bibinfo {author} {\bibfnamefont {C.}~\bibnamefont {Klewe}}, \bibinfo
  {author} {\bibfnamefont {T.}~\bibnamefont {Kuschel}}, \bibinfo {author}
  {\bibfnamefont {J.-M.}\ \bibnamefont {Schmalhorst}}, \bibinfo {author}
  {\bibfnamefont {G.}~\bibnamefont {Reiss}}, \bibinfo {author} {\bibfnamefont
  {L.}~\bibnamefont {Shen}}, \bibinfo {author} {\bibfnamefont {A.}~\bibnamefont
  {Gupta}}, \bibinfo {author} {\bibfnamefont {Y.-T.}\ \bibnamefont {Chen}},
  \bibinfo {author} {\bibfnamefont {G.~E.~W.}\ \bibnamefont {Bauer}}, \bibinfo
  {author} {\bibfnamefont {E.}~\bibnamefont {Saitoh}}, \ and\ \bibinfo {author}
  {\bibfnamefont {S.~T.~B.}\ \bibnamefont {Goennenwein}},\ }\href {\doibase
  10.1103/PhysRevB.87.224401} {\bibfield  {journal} {\bibinfo  {journal} {Phys.
  Rev. B}\ }\textbf {\bibinfo {volume} {87}},\ \bibinfo {pages} {224401}
  (\bibinfo {year} {2013})}\BibitemShut {NoStop}%
\bibitem [{\citenamefont {Sch{\"u}tz}\ \emph {et~al.}(1989)\citenamefont
  {Sch{\"u}tz}, \citenamefont {Wienke}, \citenamefont {Wilhelm}, \citenamefont
  {Wagner}, \citenamefont {Frahm},\ and\ \citenamefont {Kienle}}]{Schutz:1989}%
  \BibitemOpen
  \bibfield  {author} {\bibinfo {author} {\bibfnamefont {G.}~\bibnamefont
  {Sch{\"u}tz}}, \bibinfo {author} {\bibfnamefont {R.}~\bibnamefont {Wienke}},
  \bibinfo {author} {\bibfnamefont {W.}~\bibnamefont {Wilhelm}}, \bibinfo
  {author} {\bibfnamefont {W.}~\bibnamefont {Wagner}}, \bibinfo {author}
  {\bibfnamefont {R.}~\bibnamefont {Frahm}}, \ and\ \bibinfo {author}
  {\bibfnamefont {P.}~\bibnamefont {Kienle}},\ }\href {\doibase
  https://doi.org/10.1016/0921-4526(89)90287-1} {\bibfield  {journal} {\bibinfo
   {journal} {Physica B}\ }\textbf {\bibinfo {volume} {158}},\ \bibinfo {pages}
  {284 } (\bibinfo {year} {1989})}\BibitemShut {NoStop}%
\bibitem [{\citenamefont {Schütz}\ \emph {et~al.}(1990)\citenamefont
  {Schütz}, \citenamefont {Wienke}, \citenamefont {Wilhelm}, \citenamefont
  {Zeper}, \citenamefont {Ebert},\ and\ \citenamefont {Spörl}}]{Schutz:1990}%
  \BibitemOpen
  \bibfield  {author} {\bibinfo {author} {\bibfnamefont {G.}~\bibnamefont
  {Schütz}}, \bibinfo {author} {\bibfnamefont {R.}~\bibnamefont {Wienke}},
  \bibinfo {author} {\bibfnamefont {W.}~\bibnamefont {Wilhelm}}, \bibinfo
  {author} {\bibfnamefont {W.~B.}\ \bibnamefont {Zeper}}, \bibinfo {author}
  {\bibfnamefont {H.}~\bibnamefont {Ebert}}, \ and\ \bibinfo {author}
  {\bibfnamefont {K.}~\bibnamefont {Spörl}},\ }\href {\doibase
  10.1063/1.344903} {\bibfield  {journal} {\bibinfo  {journal} {J. Appl.
  Phys.}\ }\textbf {\bibinfo {volume} {67}},\ \bibinfo {pages} {4456} (\bibinfo
  {year} {1990})}\BibitemShut {NoStop}%
\bibitem [{\citenamefont {Maruyama}\ \emph {et~al.}(1993)\citenamefont
  {Maruyama}, \citenamefont {Koizumi}, \citenamefont {Kobayashi},\ and\
  \citenamefont {Yamazaki}}]{Maruyama:1993}%
  \BibitemOpen
  \bibfield  {author} {\bibinfo {author} {\bibfnamefont {H.}~\bibnamefont
  {Maruyama}}, \bibinfo {author} {\bibfnamefont {A.}~\bibnamefont {Koizumi}},
  \bibinfo {author} {\bibfnamefont {K.}~\bibnamefont {Kobayashi}}, \ and\
  \bibinfo {author} {\bibfnamefont {H.}~\bibnamefont {Yamazaki}},\ }\href@noop
  {} {\bibfield  {journal} {\bibinfo  {journal} {Jpn. J. Appl. Phys.}\ }\textbf
  {\bibinfo {volume} {32}},\ \bibinfo {pages} {290} (\bibinfo {year}
  {1993})}\BibitemShut {NoStop}%
\bibitem [{\citenamefont {Antel}\ \emph {et~al.}(1999)\citenamefont {Antel},
  \citenamefont {Schwickert}, \citenamefont {Lin}, \citenamefont {O'Brien},\
  and\ \citenamefont {Harp}}]{Antel:1999}%
  \BibitemOpen
  \bibfield  {author} {\bibinfo {author} {\bibfnamefont {W.~J.}\ \bibnamefont
  {Antel}}, \bibinfo {author} {\bibfnamefont {M.~M.}\ \bibnamefont
  {Schwickert}}, \bibinfo {author} {\bibfnamefont {T.}~\bibnamefont {Lin}},
  \bibinfo {author} {\bibfnamefont {W.~L.}\ \bibnamefont {O'Brien}}, \ and\
  \bibinfo {author} {\bibfnamefont {G.~R.}\ \bibnamefont {Harp}},\ }\href
  {\doibase 10.1103/PhysRevB.60.12933} {\bibfield  {journal} {\bibinfo
  {journal} {Phys. Rev. B}\ }\textbf {\bibinfo {volume} {60}},\ \bibinfo
  {pages} {12933} (\bibinfo {year} {1999})}\BibitemShut {NoStop}%
\bibitem [{\citenamefont {Wilhelm}\ \emph {et~al.}(2000)\citenamefont
  {Wilhelm}, \citenamefont {Poulopoulos}, \citenamefont {Ceballos},
  \citenamefont {Wende}, \citenamefont {Baberschke}, \citenamefont
  {Srivastava}, \citenamefont {Benea}, \citenamefont {Ebert}, \citenamefont
  {Angelakeris}, \citenamefont {Flevaris}, \citenamefont {Niarchos},
  \citenamefont {Rogalev},\ and\ \citenamefont {Brookes}}]{Wilhelm:2000}%
  \BibitemOpen
  \bibfield  {author} {\bibinfo {author} {\bibfnamefont {F.}~\bibnamefont
  {Wilhelm}}, \bibinfo {author} {\bibfnamefont {P.}~\bibnamefont
  {Poulopoulos}}, \bibinfo {author} {\bibfnamefont {G.}~\bibnamefont
  {Ceballos}}, \bibinfo {author} {\bibfnamefont {H.}~\bibnamefont {Wende}},
  \bibinfo {author} {\bibfnamefont {K.}~\bibnamefont {Baberschke}}, \bibinfo
  {author} {\bibfnamefont {P.}~\bibnamefont {Srivastava}}, \bibinfo {author}
  {\bibfnamefont {D.}~\bibnamefont {Benea}}, \bibinfo {author} {\bibfnamefont
  {H.}~\bibnamefont {Ebert}}, \bibinfo {author} {\bibfnamefont
  {M.}~\bibnamefont {Angelakeris}}, \bibinfo {author} {\bibfnamefont {N.~K.}\
  \bibnamefont {Flevaris}}, \bibinfo {author} {\bibfnamefont {D.}~\bibnamefont
  {Niarchos}}, \bibinfo {author} {\bibfnamefont {A.}~\bibnamefont {Rogalev}}, \
  and\ \bibinfo {author} {\bibfnamefont {N.~B.}\ \bibnamefont {Brookes}},\
  }\href {\doibase 10.1103/PhysRevLett.85.413} {\bibfield  {journal} {\bibinfo
  {journal} {Phys. Rev. Lett.}\ }\textbf {\bibinfo {volume} {85}},\ \bibinfo
  {pages} {413} (\bibinfo {year} {2000})}\BibitemShut {NoStop}%
\bibitem [{\citenamefont {Poulopoulos}\ \emph {et~al.}(2001)\citenamefont
  {Poulopoulos}, \citenamefont {Wilhelm}, \citenamefont {Wende}, \citenamefont
  {Ceballos}, \citenamefont {Baberschke}, \citenamefont {Benea}, \citenamefont
  {Ebert}, \citenamefont {Angelakeris}, \citenamefont {Flevaris}, \citenamefont
  {Rogalev},\ and\ \citenamefont {Brookes}}]{Poulopoulos:2001}%
  \BibitemOpen
  \bibfield  {author} {\bibinfo {author} {\bibfnamefont {P.}~\bibnamefont
  {Poulopoulos}}, \bibinfo {author} {\bibfnamefont {F.}~\bibnamefont
  {Wilhelm}}, \bibinfo {author} {\bibfnamefont {H.}~\bibnamefont {Wende}},
  \bibinfo {author} {\bibfnamefont {G.}~\bibnamefont {Ceballos}}, \bibinfo
  {author} {\bibfnamefont {K.}~\bibnamefont {Baberschke}}, \bibinfo {author}
  {\bibfnamefont {D.}~\bibnamefont {Benea}}, \bibinfo {author} {\bibfnamefont
  {H.}~\bibnamefont {Ebert}}, \bibinfo {author} {\bibfnamefont
  {M.}~\bibnamefont {Angelakeris}}, \bibinfo {author} {\bibfnamefont {N.~K.}\
  \bibnamefont {Flevaris}}, \bibinfo {author} {\bibfnamefont {A.}~\bibnamefont
  {Rogalev}}, \ and\ \bibinfo {author} {\bibfnamefont {N.~B.}\ \bibnamefont
  {Brookes}},\ }\href {\doibase 10.1063/1.1345862} {\bibfield  {journal}
  {\bibinfo  {journal} {J. Appl. Phys.}\ }\textbf {\bibinfo {volume} {89}},\
  \bibinfo {pages} {3874} (\bibinfo {year} {2001})}\BibitemShut {NoStop}%
\bibitem [{\citenamefont {Wilhelm}\ \emph {et~al.}(2003)\citenamefont
  {Wilhelm}, \citenamefont {Poulopoulos}, \citenamefont {Scherz}, \citenamefont
  {Wende}, \citenamefont {Baberschke}, \citenamefont {Angelakeris},
  \citenamefont {Flevaris}, \citenamefont {Goulon},\ and\ \citenamefont
  {Rogalev}}]{Wilhelm:2003}%
  \BibitemOpen
  \bibfield  {author} {\bibinfo {author} {\bibfnamefont {F.}~\bibnamefont
  {Wilhelm}}, \bibinfo {author} {\bibfnamefont {P.}~\bibnamefont
  {Poulopoulos}}, \bibinfo {author} {\bibfnamefont {A.}~\bibnamefont {Scherz}},
  \bibinfo {author} {\bibfnamefont {H.}~\bibnamefont {Wende}}, \bibinfo
  {author} {\bibfnamefont {K.}~\bibnamefont {Baberschke}}, \bibinfo {author}
  {\bibfnamefont {M.}~\bibnamefont {Angelakeris}}, \bibinfo {author}
  {\bibfnamefont {N.~K.}\ \bibnamefont {Flevaris}}, \bibinfo {author}
  {\bibfnamefont {J.}~\bibnamefont {Goulon}}, \ and\ \bibinfo {author}
  {\bibfnamefont {A.}~\bibnamefont {Rogalev}},\ }\href {\doibase
  10.1002/pssa.200306346} {\bibfield  {journal} {\bibinfo  {journal} {Phys.
  Status Solidi A}\ }\textbf {\bibinfo {volume} {196}},\ \bibinfo {pages} {33}
  (\bibinfo {year} {2003})}\BibitemShut {NoStop}%
\bibitem [{\citenamefont {Suzuki}\ \emph {et~al.}(2005)\citenamefont {Suzuki},
  \citenamefont {Muraoka}, \citenamefont {Inaba}, \citenamefont {Miyagawa},
  \citenamefont {Kawamura}, \citenamefont {Shimatsu}, \citenamefont {Maruyama},
  \citenamefont {Ishimatsu}, \citenamefont {Isohama},\ and\ \citenamefont
  {Sonobe}}]{Suzuki:2005}%
  \BibitemOpen
  \bibfield  {author} {\bibinfo {author} {\bibfnamefont {M.}~\bibnamefont
  {Suzuki}}, \bibinfo {author} {\bibfnamefont {H.}~\bibnamefont {Muraoka}},
  \bibinfo {author} {\bibfnamefont {Y.}~\bibnamefont {Inaba}}, \bibinfo
  {author} {\bibfnamefont {H.}~\bibnamefont {Miyagawa}}, \bibinfo {author}
  {\bibfnamefont {N.}~\bibnamefont {Kawamura}}, \bibinfo {author}
  {\bibfnamefont {T.}~\bibnamefont {Shimatsu}}, \bibinfo {author}
  {\bibfnamefont {H.}~\bibnamefont {Maruyama}}, \bibinfo {author}
  {\bibfnamefont {N.}~\bibnamefont {Ishimatsu}}, \bibinfo {author}
  {\bibfnamefont {Y.}~\bibnamefont {Isohama}}, \ and\ \bibinfo {author}
  {\bibfnamefont {Y.}~\bibnamefont {Sonobe}},\ }\href {\doibase
  10.1103/PhysRevB.72.054430} {\bibfield  {journal} {\bibinfo  {journal} {Phys.
  Rev. B}\ }\textbf {\bibinfo {volume} {72}},\ \bibinfo {pages} {054430}
  (\bibinfo {year} {2005})}\BibitemShut {NoStop}%
\bibitem [{\citenamefont {Lu}\ \emph {et~al.}(2013)\citenamefont {Lu},
  \citenamefont {Choi}, \citenamefont {Ortega}, \citenamefont {Cheng},
  \citenamefont {Cai}, \citenamefont {Huang}, \citenamefont {Sun},\ and\
  \citenamefont {Chien}}]{Lu:2013}%
  \BibitemOpen
  \bibfield  {author} {\bibinfo {author} {\bibfnamefont {Y.~M.}\ \bibnamefont
  {Lu}}, \bibinfo {author} {\bibfnamefont {Y.}~\bibnamefont {Choi}}, \bibinfo
  {author} {\bibfnamefont {C.~M.}\ \bibnamefont {Ortega}}, \bibinfo {author}
  {\bibfnamefont {X.~M.}\ \bibnamefont {Cheng}}, \bibinfo {author}
  {\bibfnamefont {J.~W.}\ \bibnamefont {Cai}}, \bibinfo {author} {\bibfnamefont
  {S.~Y.}\ \bibnamefont {Huang}}, \bibinfo {author} {\bibfnamefont
  {L.}~\bibnamefont {Sun}}, \ and\ \bibinfo {author} {\bibfnamefont {C.~L.}\
  \bibnamefont {Chien}},\ }\href {\doibase 10.1103/PhysRevLett.110.147207}
  {\bibfield  {journal} {\bibinfo  {journal} {Phys. Rev. Lett.}\ }\textbf
  {\bibinfo {volume} {110}},\ \bibinfo {pages} {147207} (\bibinfo {year}
  {2013})}\BibitemShut {NoStop}%
\bibitem [{\citenamefont {Gepr\"ags}\ \emph {et~al.}(2012)\citenamefont
  {Gepr\"ags}, \citenamefont {Meyer}, \citenamefont {Altmannshofer},
  \citenamefont {Opel}, \citenamefont {Wilhelm}, \citenamefont {Rogalev},
  \citenamefont {Gross},\ and\ \citenamefont {Goennenwein}}]{Geprags:2012}%
  \BibitemOpen
  \bibfield  {author} {\bibinfo {author} {\bibfnamefont {S.}~\bibnamefont
  {Gepr\"ags}}, \bibinfo {author} {\bibfnamefont {S.}~\bibnamefont {Meyer}},
  \bibinfo {author} {\bibfnamefont {S.}~\bibnamefont {Altmannshofer}}, \bibinfo
  {author} {\bibfnamefont {M.}~\bibnamefont {Opel}}, \bibinfo {author}
  {\bibfnamefont {F.}~\bibnamefont {Wilhelm}}, \bibinfo {author} {\bibfnamefont
  {A.}~\bibnamefont {Rogalev}}, \bibinfo {author} {\bibfnamefont
  {R.}~\bibnamefont {Gross}}, \ and\ \bibinfo {author} {\bibfnamefont
  {S.~T.~B.}\ \bibnamefont {Goennenwein}},\ }\href {\doibase 10.1063/1.4773509}
  {\bibfield  {journal} {\bibinfo  {journal} {Appl. Phys. Lett.}\ }\textbf
  {\bibinfo {volume} {101}},\ \bibinfo {pages} {262407} (\bibinfo {year}
  {2012})}\BibitemShut {NoStop}%
\bibitem [{\citenamefont {Gepr\"ags}\ \emph {et~al.}(2013)\citenamefont
  {Gepr\"ags}, \citenamefont {Goennenwein}, \citenamefont {Schneider},
  \citenamefont {Wilhelm}, \citenamefont {Ollefs}, \citenamefont {Rogalev},
  \citenamefont {Opel},\ and\ \citenamefont {Gross}}]{Geprags:2012Comment}%
  \BibitemOpen
  \bibfield  {author} {\bibinfo {author} {\bibfnamefont {S.}~\bibnamefont
  {Gepr\"ags}}, \bibinfo {author} {\bibfnamefont {S.~T.~B.}\ \bibnamefont
  {Goennenwein}}, \bibinfo {author} {\bibfnamefont {M.}~\bibnamefont
  {Schneider}}, \bibinfo {author} {\bibfnamefont {F.}~\bibnamefont {Wilhelm}},
  \bibinfo {author} {\bibfnamefont {K.}~\bibnamefont {Ollefs}}, \bibinfo
  {author} {\bibfnamefont {A.}~\bibnamefont {Rogalev}}, \bibinfo {author}
  {\bibfnamefont {M.}~\bibnamefont {Opel}}, \ and\ \bibinfo {author}
  {\bibfnamefont {R.}~\bibnamefont {Gross}},\ }\href@noop {} {\bibfield
  {journal} {\bibinfo  {journal} {arXiv:1307.4869}\ } (\bibinfo {year}
  {2013})}\BibitemShut {NoStop}%
\bibitem [{\citenamefont {Valvidares}\ \emph {et~al.}(2016)\citenamefont
  {Valvidares}, \citenamefont {Dix}, \citenamefont {Isasa}, \citenamefont
  {Ollefs}, \citenamefont {Wilhelm}, \citenamefont {Rogalev}, \citenamefont
  {S\'anchez}, \citenamefont {Pellegrin}, \citenamefont {Bedoya-Pinto},
  \citenamefont {Gargiani}, \citenamefont {Hueso}, \citenamefont {Casanova},\
  and\ \citenamefont {Fontcuberta}}]{Valvidares:2016}%
  \BibitemOpen
  \bibfield  {author} {\bibinfo {author} {\bibfnamefont {M.}~\bibnamefont
  {Valvidares}}, \bibinfo {author} {\bibfnamefont {N.}~\bibnamefont {Dix}},
  \bibinfo {author} {\bibfnamefont {M.}~\bibnamefont {Isasa}}, \bibinfo
  {author} {\bibfnamefont {K.}~\bibnamefont {Ollefs}}, \bibinfo {author}
  {\bibfnamefont {F.}~\bibnamefont {Wilhelm}}, \bibinfo {author} {\bibfnamefont
  {A.}~\bibnamefont {Rogalev}}, \bibinfo {author} {\bibfnamefont
  {F.}~\bibnamefont {S\'anchez}}, \bibinfo {author} {\bibfnamefont
  {E.}~\bibnamefont {Pellegrin}}, \bibinfo {author} {\bibfnamefont
  {A.}~\bibnamefont {Bedoya-Pinto}}, \bibinfo {author} {\bibfnamefont
  {P.}~\bibnamefont {Gargiani}}, \bibinfo {author} {\bibfnamefont {L.~E.}\
  \bibnamefont {Hueso}}, \bibinfo {author} {\bibfnamefont {F.}~\bibnamefont
  {Casanova}}, \ and\ \bibinfo {author} {\bibfnamefont {J.}~\bibnamefont
  {Fontcuberta}},\ }\href {\doibase 10.1103/PhysRevB.93.214415} {\bibfield
  {journal} {\bibinfo  {journal} {Phys. Rev. B}\ }\textbf {\bibinfo {volume}
  {93}},\ \bibinfo {pages} {214415} (\bibinfo {year} {2016})}\BibitemShut
  {NoStop}%
\bibitem [{\citenamefont {Collet}\ \emph {et~al.}(2017)\citenamefont {Collet},
  \citenamefont {Mattana}, \citenamefont {Moussy}, \citenamefont {Ollefs},
  \citenamefont {Collin}, \citenamefont {Deranlot}, \citenamefont {Anane},
  \citenamefont {Cros}, \citenamefont {Petroff}, \citenamefont {Wilhelm},\ and\
  \citenamefont {Rogalev}}]{Collet:2017}%
  \BibitemOpen
  \bibfield  {author} {\bibinfo {author} {\bibfnamefont {M.}~\bibnamefont
  {Collet}}, \bibinfo {author} {\bibfnamefont {R.}~\bibnamefont {Mattana}},
  \bibinfo {author} {\bibfnamefont {J.-B.}\ \bibnamefont {Moussy}}, \bibinfo
  {author} {\bibfnamefont {K.}~\bibnamefont {Ollefs}}, \bibinfo {author}
  {\bibfnamefont {S.}~\bibnamefont {Collin}}, \bibinfo {author} {\bibfnamefont
  {C.}~\bibnamefont {Deranlot}}, \bibinfo {author} {\bibfnamefont
  {A.}~\bibnamefont {Anane}}, \bibinfo {author} {\bibfnamefont
  {V.}~\bibnamefont {Cros}}, \bibinfo {author} {\bibfnamefont {F.}~\bibnamefont
  {Petroff}}, \bibinfo {author} {\bibfnamefont {F.}~\bibnamefont {Wilhelm}}, \
  and\ \bibinfo {author} {\bibfnamefont {A.}~\bibnamefont {Rogalev}},\ }\href
  {\doibase 10.1063/1.4987145} {\bibfield  {journal} {\bibinfo  {journal}
  {Appl. Phys. Lett.}\ }\textbf {\bibinfo {volume} {111}},\ \bibinfo {pages}
  {202401} (\bibinfo {year} {2017})}\BibitemShut {NoStop}%
\bibitem [{\citenamefont {Kikkawa}\ \emph {et~al.}(2017)\citenamefont
  {Kikkawa}, \citenamefont {Suzuki}, \citenamefont {Okabayashi}, \citenamefont
  {Uchida}, \citenamefont {Kikuchi}, \citenamefont {Qiu},\ and\ \citenamefont
  {Saitoh}}]{Kikkawa:2017}%
  \BibitemOpen
  \bibfield  {author} {\bibinfo {author} {\bibfnamefont {T.}~\bibnamefont
  {Kikkawa}}, \bibinfo {author} {\bibfnamefont {M.}~\bibnamefont {Suzuki}},
  \bibinfo {author} {\bibfnamefont {J.}~\bibnamefont {Okabayashi}}, \bibinfo
  {author} {\bibfnamefont {K.-i.}\ \bibnamefont {Uchida}}, \bibinfo {author}
  {\bibfnamefont {D.}~\bibnamefont {Kikuchi}}, \bibinfo {author} {\bibfnamefont
  {Z.}~\bibnamefont {Qiu}}, \ and\ \bibinfo {author} {\bibfnamefont
  {E.}~\bibnamefont {Saitoh}},\ }\href {\doibase 10.1103/PhysRevB.95.214416}
  {\bibfield  {journal} {\bibinfo  {journal} {Phys. Rev. B}\ }\textbf {\bibinfo
  {volume} {95}},\ \bibinfo {pages} {214416} (\bibinfo {year}
  {2017})}\BibitemShut {NoStop}%
\bibitem [{\citenamefont {Macke}\ and\ \citenamefont
  {Goering}(2014)}]{Macke:2014}%
  \BibitemOpen
  \bibfield  {author} {\bibinfo {author} {\bibfnamefont {S.}~\bibnamefont
  {Macke}}\ and\ \bibinfo {author} {\bibfnamefont {E.}~\bibnamefont
  {Goering}},\ }\href@noop {} {\bibfield  {journal} {\bibinfo  {journal} {J.
  Phys.: Condens. Matter}\ }\textbf {\bibinfo {volume} {26}},\ \bibinfo {pages}
  {363201} (\bibinfo {year} {2014})}\BibitemShut {NoStop}%
\bibitem [{\citenamefont {Kuschel}\ \emph {et~al.}(2015)\citenamefont
  {Kuschel}, \citenamefont {Klewe}, \citenamefont {Schmalhorst}, \citenamefont
  {Bertram}, \citenamefont {Kuschel}, \citenamefont {Schemme}, \citenamefont
  {Wollschl\"ager}, \citenamefont {Francoual}, \citenamefont {Strempfer},
  \citenamefont {Gupta}, \citenamefont {Meinert}, \citenamefont {G\"otz},
  \citenamefont {Meier},\ and\ \citenamefont {Reiss}}]{Kuschel:2015}%
  \BibitemOpen
  \bibfield  {author} {\bibinfo {author} {\bibfnamefont {T.}~\bibnamefont
  {Kuschel}}, \bibinfo {author} {\bibfnamefont {C.}~\bibnamefont {Klewe}},
  \bibinfo {author} {\bibfnamefont {J.-M.}\ \bibnamefont {Schmalhorst}},
  \bibinfo {author} {\bibfnamefont {F.}~\bibnamefont {Bertram}}, \bibinfo
  {author} {\bibfnamefont {O.}~\bibnamefont {Kuschel}}, \bibinfo {author}
  {\bibfnamefont {T.}~\bibnamefont {Schemme}}, \bibinfo {author} {\bibfnamefont
  {J.}~\bibnamefont {Wollschl\"ager}}, \bibinfo {author} {\bibfnamefont
  {S.}~\bibnamefont {Francoual}}, \bibinfo {author} {\bibfnamefont
  {J.}~\bibnamefont {Strempfer}}, \bibinfo {author} {\bibfnamefont
  {A.}~\bibnamefont {Gupta}}, \bibinfo {author} {\bibfnamefont
  {M.}~\bibnamefont {Meinert}}, \bibinfo {author} {\bibfnamefont
  {G.}~\bibnamefont {G\"otz}}, \bibinfo {author} {\bibfnamefont
  {D.}~\bibnamefont {Meier}}, \ and\ \bibinfo {author} {\bibfnamefont
  {G.}~\bibnamefont {Reiss}},\ }\href {\doibase 10.1103/PhysRevLett.115.097401}
  {\bibfield  {journal} {\bibinfo  {journal} {Phys. Rev. Lett.}\ }\textbf
  {\bibinfo {volume} {115}},\ \bibinfo {pages} {097401} (\bibinfo {year}
  {2015})}\BibitemShut {NoStop}%
\bibitem [{\citenamefont {Kuschel}\ \emph {et~al.}(2016)\citenamefont
  {Kuschel}, \citenamefont {Klewe}, \citenamefont {Bougiatioti}, \citenamefont
  {Kuschel}, \citenamefont {Wollschläger}, \citenamefont {Bouchenoire},
  \citenamefont {Brown}, \citenamefont {Schmalhorst}, \citenamefont {Meier},\
  and\ \citenamefont {Reiss}}]{Kuschel:2016}%
  \BibitemOpen
  \bibfield  {author} {\bibinfo {author} {\bibfnamefont {T.}~\bibnamefont
  {Kuschel}}, \bibinfo {author} {\bibfnamefont {C.}~\bibnamefont {Klewe}},
  \bibinfo {author} {\bibfnamefont {P.}~\bibnamefont {Bougiatioti}}, \bibinfo
  {author} {\bibfnamefont {O.}~\bibnamefont {Kuschel}}, \bibinfo {author}
  {\bibfnamefont {J.}~\bibnamefont {Wollschläger}}, \bibinfo {author}
  {\bibfnamefont {L.}~\bibnamefont {Bouchenoire}}, \bibinfo {author}
  {\bibfnamefont {S.~D.}\ \bibnamefont {Brown}}, \bibinfo {author}
  {\bibfnamefont {J.~M.}\ \bibnamefont {Schmalhorst}}, \bibinfo {author}
  {\bibfnamefont {D.}~\bibnamefont {Meier}}, \ and\ \bibinfo {author}
  {\bibfnamefont {G.}~\bibnamefont {Reiss}},\ }\href {\doibase
  10.1109/TMAG.2015.2512040} {\bibfield  {journal} {\bibinfo  {journal} {IEEE
  Trans. Magn.}\ }\textbf {\bibinfo {volume} {52}},\ \bibinfo {pages} {4500104}
  (\bibinfo {year} {2016})}\BibitemShut {NoStop}%
\bibitem [{\citenamefont {Bougiatioti}\ \emph
  {et~al.}(2017{\natexlab{b}})\citenamefont {Bougiatioti}, \citenamefont
  {Manos}, \citenamefont {Klewe}, \citenamefont {Meier}, \citenamefont
  {Teichert}, \citenamefont {Schmalhorst}, \citenamefont {Kuschel},\ and\
  \citenamefont {Reiss}}]{BougiatiotiJAP:2017}%
  \BibitemOpen
  \bibfield  {author} {\bibinfo {author} {\bibfnamefont {P.}~\bibnamefont
  {Bougiatioti}}, \bibinfo {author} {\bibfnamefont {O.}~\bibnamefont {Manos}},
  \bibinfo {author} {\bibfnamefont {C.}~\bibnamefont {Klewe}}, \bibinfo
  {author} {\bibfnamefont {D.}~\bibnamefont {Meier}}, \bibinfo {author}
  {\bibfnamefont {N.}~\bibnamefont {Teichert}}, \bibinfo {author}
  {\bibfnamefont {J.-M.}\ \bibnamefont {Schmalhorst}}, \bibinfo {author}
  {\bibfnamefont {T.}~\bibnamefont {Kuschel}}, \ and\ \bibinfo {author}
  {\bibfnamefont {G.}~\bibnamefont {Reiss}},\ }\href@noop {} {\bibfield
  {journal} {\bibinfo  {journal} {J. Appl. Phys.}\ }\textbf {\bibinfo {volume}
  {122}},\ \bibinfo {pages} {225101} (\bibinfo {year}
  {2017}{\natexlab{b}})}\BibitemShut {NoStop}%
\bibitem [{\citenamefont {Klewe}\ \emph {et~al.}(2016)\citenamefont {Klewe},
  \citenamefont {Kuschel}, \citenamefont {Schmalhorst}, \citenamefont
  {Bertram}, \citenamefont {Kuschel}, \citenamefont {Wollschl\"ager},
  \citenamefont {Strempfer}, \citenamefont {Meinert},\ and\ \citenamefont
  {Reiss}}]{Klewe:2016}%
  \BibitemOpen
  \bibfield  {author} {\bibinfo {author} {\bibfnamefont {C.}~\bibnamefont
  {Klewe}}, \bibinfo {author} {\bibfnamefont {T.}~\bibnamefont {Kuschel}},
  \bibinfo {author} {\bibfnamefont {J.-M.}\ \bibnamefont {Schmalhorst}},
  \bibinfo {author} {\bibfnamefont {F.}~\bibnamefont {Bertram}}, \bibinfo
  {author} {\bibfnamefont {O.}~\bibnamefont {Kuschel}}, \bibinfo {author}
  {\bibfnamefont {J.}~\bibnamefont {Wollschl\"ager}}, \bibinfo {author}
  {\bibfnamefont {J.}~\bibnamefont {Strempfer}}, \bibinfo {author}
  {\bibfnamefont {M.}~\bibnamefont {Meinert}}, \ and\ \bibinfo {author}
  {\bibfnamefont {G.}~\bibnamefont {Reiss}},\ }\href {\doibase
  10.1103/PhysRevB.93.214440} {\bibfield  {journal} {\bibinfo  {journal} {Phys.
  Rev. B}\ }\textbf {\bibinfo {volume} {93}},\ \bibinfo {pages} {214440}
  (\bibinfo {year} {2016})}\BibitemShut {NoStop}%
\bibitem [{\citenamefont {Strempfer}\ \emph {et~al.}(2013)\citenamefont
  {Strempfer}, \citenamefont {Francoual}, \citenamefont {Reuther},
  \citenamefont {Shukla}, \citenamefont {Skaugen}, \citenamefont
  {Schulte-Schrepping}, \citenamefont {Kracht},\ and\ \citenamefont
  {Franz}}]{Strempfer:2013}%
  \BibitemOpen
  \bibfield  {author} {\bibinfo {author} {\bibfnamefont {J.}~\bibnamefont
  {Strempfer}}, \bibinfo {author} {\bibfnamefont {S.}~\bibnamefont
  {Francoual}}, \bibinfo {author} {\bibfnamefont {D.}~\bibnamefont {Reuther}},
  \bibinfo {author} {\bibfnamefont {D.~K.}\ \bibnamefont {Shukla}}, \bibinfo
  {author} {\bibfnamefont {A.}~\bibnamefont {Skaugen}}, \bibinfo {author}
  {\bibfnamefont {H.}~\bibnamefont {Schulte-Schrepping}}, \bibinfo {author}
  {\bibfnamefont {T.}~\bibnamefont {Kracht}}, \ and\ \bibinfo {author}
  {\bibfnamefont {H.}~\bibnamefont {Franz}},\ }\href@noop {} {\bibfield
  {journal} {\bibinfo  {journal} {J. Synchrotron Radiat.}\ }\textbf {\bibinfo
  {volume} {20}},\ \bibinfo {pages} {541} (\bibinfo {year} {2013})}\BibitemShut
  {NoStop}%
\bibitem [{SM()}]{SM}%
  \BibitemOpen
  \href@noop {} {\bibinfo  {journal} {See Supplemental Material at [URL] for
  additional information about the XRMR technique and the extracted magnetic
  moments}\ }\BibitemShut {NoStop}%
\bibitem [{\citenamefont {Geissler}\ \emph {et~al.}(2001)\citenamefont
  {Geissler}, \citenamefont {Goering}, \citenamefont {Justen}, \citenamefont
  {Weigand}, \citenamefont {Sch\"utz}, \citenamefont {Langer}, \citenamefont
  {Schmitz}, \citenamefont {Maletta},\ and\ \citenamefont
  {Mattheis}}]{Geissler:2001}%
  \BibitemOpen
\bibfield  {journal} {  }\bibfield  {author} {\bibinfo {author} {\bibfnamefont
  {J.}~\bibnamefont {Geissler}}, \bibinfo {author} {\bibfnamefont
  {E.}~\bibnamefont {Goering}}, \bibinfo {author} {\bibfnamefont
  {M.}~\bibnamefont {Justen}}, \bibinfo {author} {\bibfnamefont
  {F.}~\bibnamefont {Weigand}}, \bibinfo {author} {\bibfnamefont
  {G.}~\bibnamefont {Sch\"utz}}, \bibinfo {author} {\bibfnamefont
  {J.}~\bibnamefont {Langer}}, \bibinfo {author} {\bibfnamefont
  {D.}~\bibnamefont {Schmitz}}, \bibinfo {author} {\bibfnamefont
  {H.}~\bibnamefont {Maletta}}, \ and\ \bibinfo {author} {\bibfnamefont
  {R.}~\bibnamefont {Mattheis}},\ }\href {\doibase 10.1103/PhysRevB.65.020405}
  {\bibfield  {journal} {\bibinfo  {journal} {Phys. Rev. B}\ }\textbf {\bibinfo
  {volume} {65}},\ \bibinfo {pages} {020405} (\bibinfo {year}
  {2001})}\BibitemShut {NoStop}%
\bibitem [{\citenamefont {Bonnenberg}\ \emph {et~al.}(1986)\citenamefont
  {Bonnenberg}, \citenamefont {Hempel},\ and\ \citenamefont
  {Wijn}}]{LandoltBornstein1986}%
  \BibitemOpen
  \bibfield  {author} {\bibinfo {author} {\bibfnamefont {D.}~\bibnamefont
  {Bonnenberg}}, \bibinfo {author} {\bibfnamefont {K.~A.}\ \bibnamefont
  {Hempel}}, \ and\ \bibinfo {author} {\bibfnamefont {H.}~\bibnamefont
  {Wijn}},\ }\href@noop {} {\emph {\bibinfo {title} {Atomic magnetic moment,
  magnetic moment density, g and g' factor}}}\ (\bibinfo  {publisher}
  {Springer-Verlag Berlin Heidelberg},\ \bibinfo {year} {1986})\BibitemShut
  {NoStop}%
\bibitem [{\citenamefont {Bozorth}(1951)}]{Bozorth:1951}%
  \BibitemOpen
  \bibfield  {author} {\bibinfo {author} {\bibfnamefont {D.~M.}\ \bibnamefont
  {Bozorth}},\ }\href@noop {} {\emph {\bibinfo {title} {{Ferromagnetism}}}}\
  (\bibinfo  {publisher} {Van Nostrand, New York},\ \bibinfo {year}
  {1951})\BibitemShut {NoStop}%
\bibitem [{\citenamefont {{J.L Costa-Kr\"amer and J.L Men\'endez and A
  Cebollada and F Briones and D Garci\'{a} and A
  Hernando}}(2000)}]{Costa:2000}%
  \BibitemOpen
  \bibfield  {author} {\bibinfo {author} {\bibnamefont {{J.L Costa-Kr\"amer and
  J.L Men\'endez and A Cebollada and F Briones and D Garci\'{a} and A
  Hernando}}},\ }\href@noop {} {\bibfield  {journal} {\bibinfo  {journal} {J.
  Magn. Magn. Mater.}\ }\textbf {\bibinfo {volume} {210}},\ \bibinfo {pages}
  {341 } (\bibinfo {year} {2000})}\BibitemShut {NoStop}%
\bibitem [{\citenamefont {{T Kuschel and H Bardenhagen and H Wilkens and R
  Schubert and J Hamrle and J Pi\v stora and J
  Wollschl\"ager}}(2011)}]{Kuschel:2011}%
  \BibitemOpen
  \bibfield  {author} {\bibinfo {author} {\bibnamefont {{T Kuschel and H
  Bardenhagen and H Wilkens and R Schubert and J Hamrle and J Pi\v stora and J
  Wollschl\"ager}}},\ }\href@noop {} {\bibfield  {journal} {\bibinfo  {journal}
  {J. Phys. D: Appl. Phys.}\ }\textbf {\bibinfo {volume} {44}},\ \bibinfo
  {pages} {265003} (\bibinfo {year} {2011})}\BibitemShut {NoStop}%
\bibitem [{\citenamefont {Shikada}\ \emph {et~al.}(2009)\citenamefont
  {Shikada}, \citenamefont {Ohtake}, \citenamefont {Kirino},\ and\
  \citenamefont {Futamoto}}]{Shikada:2009}%
  \BibitemOpen
  \bibfield  {author} {\bibinfo {author} {\bibfnamefont {K.}~\bibnamefont
  {Shikada}}, \bibinfo {author} {\bibfnamefont {M.}~\bibnamefont {Ohtake}},
  \bibinfo {author} {\bibfnamefont {F.}~\bibnamefont {Kirino}}, \ and\ \bibinfo
  {author} {\bibfnamefont {M.}~\bibnamefont {Futamoto}},\ }\href {\doibase
  10.1063/1.3067854} {\bibfield  {journal} {\bibinfo  {journal} {J. Appl.
  Phys.}\ }\textbf {\bibinfo {volume} {105}},\ \bibinfo {pages} {07C303}
  (\bibinfo {year} {2009})}\BibitemShut {NoStop}%
\bibitem [{\citenamefont {{T Kuschel and J Hamrle and J Pištora and K Saito
  and S Bosu and Y Sakuraba and K Takanashi and J
  Wollschl\"ager}}(2012)}]{Kuschel:2012}%
  \BibitemOpen
  \bibfield  {author} {\bibinfo {author} {\bibnamefont {{T Kuschel and J Hamrle
  and J Pištora and K Saito and S Bosu and Y Sakuraba and K Takanashi and J
  Wollschl\"ager}}},\ }\href@noop {} {\bibfield  {journal} {\bibinfo  {journal}
  {J. Phys. D: Appl. Phys.}\ }\textbf {\bibinfo {volume} {45}},\ \bibinfo
  {pages} {495002} (\bibinfo {year} {2012})}\BibitemShut {NoStop}%
\end{thebibliography}%

\end{document}